\documentclass[aps,prb,twocolumn,floatfix,superscriptaddress]{revtex4-2}

\usepackage{amsmath}
\usepackage{amssymb}
\usepackage{amsfonts}
\usepackage{graphicx}
\usepackage{bm}
\usepackage{hyperref}
\usepackage{xcolor}
\usepackage{booktabs}
\usepackage{braket}

\hypersetup{
    colorlinks=true,
    linkcolor=blue!60!black,
    citecolor=blue!60!black,
    urlcolor=blue!60!black
}

\newcommand{\Heph}{\hat{H}_{\mathrm{e-ph}}}
\newcommand{\Fcal}{\mathcal{F}}

\newcommand{\Vq}{V_{\mathbf{q}}}
\newcommand{\alphah}{\alpha_h}
\newcommand{\alphae}{\alpha_e}
\newcommand{\qmax}{q_{\mathrm{max}}}
\newcommand{\alat}{a_{\mathrm{lat}}}
\newcommand{\aX}{a_X}

\begin{document}
\title{Parity-controlled electron-hole interference in exciton-phonon coupling}

\author{Michael O. Atambo}
\email{michael.atambo@tukenya.ac.ke}
\affiliation{Department of Physics, Earth and Environmental Science, Technical University of Kenya, Nairobi, Kenya}

\date{\today}

\begin{abstract}
We demonstrate that exciton-phonon coupling in polar semiconductors
is governed by a parity-controlled interference selection rule. By
performing an exact multipole expansion of the excitonic form factor
and validating it against numerical integration of hydrogenic
envelope functions, we show that the long-range infrared divergence
of the Fr\"ohlich interaction is exactly canceled for elastic
scattering between excitonic states of the same parity. The
ground-state exciton is thereby protected from long-wavelength polar
phonons by destructive electron-hole interference. In contrast,
transitions between states of opposite parity exhibit constructive
interference, preserving a finite, robust coupling to macroscopic
polar fields independent of band-structure details. Mass asymmetry
between the electron and hole activates higher-order multipole terms
in the elastic channel but leaves the constructive inelastic channel
essentially unaffected. The selection rule is dimensionally invariant,
applying to bulk and two-dimensional systems alike, and naturally
explains the anomalously weak phonon dressing observed in halide
perovskites as well as the strong phonon sidebands in
transition-metal dichalcogenides. Our framework provides a broad,
analytically exact criterion for exciton-phonon coupling strength,
offering a design principle for engineering excitonic materials with
tailored phonon interactions.
\end{abstract}


\maketitle

\section{\label{sec:intro}Introduction}

Exciton--phonon coupling fundamentally dictates the finite-temperature
optical properties of semiconductors, governing phenomena such as
phonon-assisted absorption, photoluminescence linewidths, and the
formation of exciton-polarons~\cite{giustino_rmp_2017}. While the
coupling of single charge carriers to lattice vibrations is
well-understood---most notably through the Fr\"ohlich interaction
with longitudinal optical (LO) phonons---the coupling of composite,
neutral excitons to these same macroscopic fields remains a subject
of conceptual subtlety.

The bare Fr\"ohlich interaction is characterized by a long-range
macroscopic electric field that diverges in the long-wavelength limit
as $1/q$. For single polarons, this infrared divergence leads to
strong lattice dressing and significant mass renormalization. However,
an exciton is an electrically neutral bound state of an electron and
a hole. A naive application of the independent-particle picture
suggests that the strong, divergent couplings of the individual
constituents should simply add up, resulting in a heavily dressed
exciton. Yet, physical intuition and empirical observations in
various material classes suggest that neutral excitons can remain
remarkably robust against polar phonon
scattering~\cite{ Diab2016JPCLETT,Wright2016,giustino_rmp_2017}.

In this work, we resolve this apparent paradox by isolating the exact
exciton--phonon vertex. We demonstrate that electron--hole
interference in exciton--phonon coupling is not merely a quantitative
reduction, but is governed by a strict \emph{parity-controlled
selection rule}. By performing an exact multipole expansion of the
excitonic form factor, we show that the long-range $1/q$ divergence
of the Fr\"ohlich potential is perfectly cured by the internal parity
of the excitonic transition. Specifically, elastic scattering channels
(where the initial and final exciton states share the same parity,
such as $1s \to 1s$) exhibit strictly destructive interference in
the long-wavelength limit, rendering the ground-state exciton immune
to infrared Fr\"ohlich divergence. Conversely, inelastic channels
connecting states of opposite parity (such as $1s \to 2p$, or
interlayer excitons with broken spatial inversion symmetry) exhibit
constructive interference, preserving a finite, macroscopic coupling
to long-wavelength polar phonons.

Furthermore, we establish that mass asymmetry between the electron and
the hole acts as a control knob for the elastic channel. While finite
mass asymmetry allows higher-order multipole terms (quadrupolar and
beyond) to break the perfect cancellation, it has virtually no effect
on the constructive inelastic channels. These analytical results are
validated by exact numerical integration of hydrogenic envelope
functions, culminating in a unifying phase diagram that maps the
integrated polaronic dressing parameter across the space of exciton
size and mass asymmetry. Our framework provides a computationally
inexpensive, analytically exact foundation for understanding
exciton--phonon interactions, moving beyond the need for exhaustive
first-principles supercell calculations to capture emergent correlated
many-body effects.

\medskip
\noindent\textit{Remark (relation to concurrent lattice studies).}-The elastic
coefficient derived here, $\tfrac12\eta\,q^2\langle z^2\rangle$, is the
frozen-envelope (parabolic-continuum) value, exact by Galilean
invariance for a $K$-independent envelope. The lattice study of Ref.~\cite{atambo_lattice_2026} reports
$\tfrac32\eta' q^2\langle r^2\rangle$ with
$\eta'=(t_e-t_h)/(t_e+t_h)$; the additional $\eta'$ is the
center-of-mass phase-twist (recoil) contribution present in the lattice
COM gauge and absent here by construction. The two coefficients are
complementary partitions of the same activation physics, not
inconsistent results. The leading-order rule proved here is extended to an exact finite-$q$ identity in~\cite{atambo_recoil_2026}; in addition, the present work provides the inelastic multipole expansion, the
$(\Lambda,\eta)$ dressing phase diagram, and the 2D, interlayer, and charge-transfer extensions.

\section{\label{sec:model}Theoretical Model}

\subsection{\label{sec:hamiltonian}The Exciton--Phonon Hamiltonian}

We begin with the standard Fr\"ohlich Hamiltonian describing the
interaction between charge carriers and the macroscopic longitudinal
optical (LO) phonon field in a polar crystal. The interaction
Hamiltonian in momentum space is given by:
\begin{equation}
\label{eq:hamiltonian}
\Heph = \sum_{\mathbf{q}} \Vq\, \hat{\rho}(\mathbf{q})
\left(b_{\mathbf{q}} + b_{-\mathbf{q}}^\dagger\right),
\end{equation}
where $b_{\mathbf{q}}^\dagger$ and $b_{\mathbf{q}}$ are the phonon
creation and annihilation operators, and $\Vq \propto 1/q$ is the
bare Fr\"ohlich coupling potential. The charge density operator
$\hat{\rho}(\mathbf{q})$ for the electron ($e$) and hole ($h$) is:
\begin{equation}
\label{eq:charge_density}
\hat{\rho}(\mathbf{q})
= -e^{i\mathbf{q}\cdot\mathbf{r}_e}
+ e^{i\mathbf{q}\cdot\mathbf{r}_h},
\end{equation}
where we have factored out the elementary charge $e$.

\subsection{\label{sec:coordinates}Center-of-Mass Transformation}

To evaluate the matrix elements between excitonic eigenstates, we
transform from the individual particle coordinates
$(\mathbf{r}_e, \mathbf{r}_h)$ to the center-of-mass (CM) coordinate
$\mathbf{R}$ and the relative coordinate $\mathbf{r}$:
\begin{equation}
\label{eq:coordinates}
\mathbf{R} = \frac{m_e \mathbf{r}_e + m_h \mathbf{r}_h}{M},
\qquad
\mathbf{r} = \mathbf{r}_e - \mathbf{r}_h,
\end{equation}
where $M = m_e + m_h$ is the total exciton mass. Inverting these
relations yields:
\begin{equation}
\label{eq:inverse_coords}
\mathbf{r}_e = \mathbf{R} + \alphah\, \mathbf{r},
\qquad
\mathbf{r}_h = \mathbf{R} - \alphae\, \mathbf{r},
\end{equation}
where the mass fractions are defined as $\alphae = m_e/M$ and
$\alphah = m_h/M$, satisfying $\alphae + \alphah = 1$. We introduce
the mass asymmetry parameter:
\begin{equation}
\label{eq:eta}
\eta = \frac{m_h - m_e}{m_h + m_e},
\end{equation}
such that $\alphah = (1+\eta)/2$ and $\alphae = (1-\eta)/2$.

Substituting Eq.~\eqref{eq:inverse_coords} into the charge density
operator, the CM dependence factors out entirely as
$e^{i\mathbf{q}\cdot\mathbf{R}}$, enforcing crystal momentum
conservation ($\mathbf{Q}' = \mathbf{Q} + \mathbf{q}$). The
remaining internal structure of the exciton is captured by the
excitonic form factor, $\Fcal_{n'n}(\mathbf{q})$:
\begin{equation}
\label{eq:form_factor}
\Fcal_{n'n}(\mathbf{q})
= \langle \Phi_{n'} |
\left( e^{-i\alphae \mathbf{q}\cdot\mathbf{r}}
     - e^{i\alphah \mathbf{q}\cdot\mathbf{r}} \right)
| \Phi_n \rangle,
\end{equation}
where $|\Phi_n\rangle$ represents the internal envelope wavefunction
of the exciton (e.g., the $1s$ or $2p$ hydrogenic states).

\subsection{\label{sec:multipole}Multipole Expansion and the Parity
Selection Rule}

To understand the long-wavelength limit ($q \to 0$), we Taylor expand
the exponential operators inside the matrix element:
\begin{align}
\label{eq:taylor}
&e^{-i\alphae \mathbf{q}\cdot\mathbf{r}}
- e^{i\alphah \mathbf{q}\cdot\mathbf{r}} \nonumber\\
&= \left(1 - i\alphae \mathbf{q}\cdot\mathbf{r}
   - \tfrac{1}{2}\alphae^2 (\mathbf{q}\cdot\mathbf{r})^2 + \cdots
   \right) \nonumber\\
&\quad - \left(1 + i\alphah \mathbf{q}\cdot\mathbf{r}
   - \tfrac{1}{2}\alphah^2 (\mathbf{q}\cdot\mathbf{r})^2 + \cdots
   \right).
\end{align}
Grouping terms by the order of $q$ reveals the underlying multipole
physics:
\begin{align}
\text{0th order (Monopole)} &: \quad -1 + 1 = 0, \label{eq:monopole}\\
\text{1st order (Dipole)} &: \quad -i(\alphae + \alphah)\,\mathbf{q}\cdot\mathbf{r}
= -i\,\mathbf{q}\cdot\mathbf{r}, \label{eq:dipole}\\
\text{2nd order (Quadrupole)} &: \quad
\tfrac{1}{2}(\alphah^2 - \alphae^2)(\mathbf{q}\cdot\mathbf{r})^2
= \tfrac{1}{2}\eta\,(\mathbf{q}\cdot\mathbf{r})^2.
\label{eq:quadrupole}
\end{align}
The 0th-order term vanishes identically, reflecting the exact
neutrality of the exciton.

The behavior of the remaining terms depends entirely on the parity of
the initial and final exciton states. For an elastic transition
between states of the same parity (e.g., $1s \to 1s$), the dipole
operator $\mathbf{r}$ (which is odd under spatial inversion) yields a
strictly zero expectation value:
$\langle 1s | \mathbf{r} | 1s \rangle = 0$.
Consequently, the leading non-vanishing term is the quadrupole,
scaling as $\mathcal{O}(q^2)$. Because the effective scattering
matrix element squared is:
\begin{equation}
\label{eq:g_squared}
|g_X(\mathbf{q})|^2
= |\Vq|^2\, |\Fcal(\mathbf{q})|^2
\propto \frac{1}{q^2}\, |\Fcal(\mathbf{q})|^2,
\end{equation}
the elastic coupling scales as $(1/q^2) \times q^4 = q^2$. In the
$q \to 0$ limit, the elastic coupling vanishes entirely. The
ground-state exciton is perfectly protected from the infrared
divergence of the Fr\"ohlich field.

Conversely, for an inelastic transition between states of opposite
parity (e.g., $1s \to 2p$), the dipole expectation value is non-zero.
The form factor scales as $\mathcal{O}(q^1)$, and the effective
coupling scales as $(1/q^2) \times q^2 = q^0$. The coupling remains
finite and macroscopic in the long-wavelength limit. This establishes
a fundamental dichotomy:

\begin{equation}
\label{eq:selection_rule}
\boxed{
\Fcal_{n'n}(q) =
\begin{cases}
g(\alphae q) - g(\alphah q) & \text{same parity } (s\!\to\!s) \\
A(\alphah q) + A(\alphae q) & \text{opposite parity } (s\!\to\!p)
\end{cases}
}
\end{equation}

\emph{Same-parity transitions are protected by destructive
interference, while opposite-parity transitions are enhanced by
constructive interference.}

\section{\label{sec:results}Results and Discussion}

\subsection{\label{sec:validation}Exact Validation of the Asymptotic
Limits}

To validate the multipole expansion derived in
Sec.~\ref{sec:multipole}, we perform exact numerical integration of
the excitonic form factors using the full 3D hydrogenic envelope
functions. The $1s$ ground-state envelope is
$\Phi_{1s}(\mathbf{r}) = (\pi \aX^3)^{-1/2} e^{-r/\aX}$, and the
$2p$ excited-state envelope is
$\Phi_{2p_z}(\mathbf{r}) = (32\pi \aX^5)^{-1/2}\, r\, e^{-r/2\aX}
\cos\theta$, where $\aX$ is the exciton Bohr radius. The form factors
$\Fcal_{n'n}(\mathbf{q})$ are computed via direct numerical
integration over all three spatial dimensions.

\begin{figure}[b]
\centering
\includegraphics[width=\columnwidth]{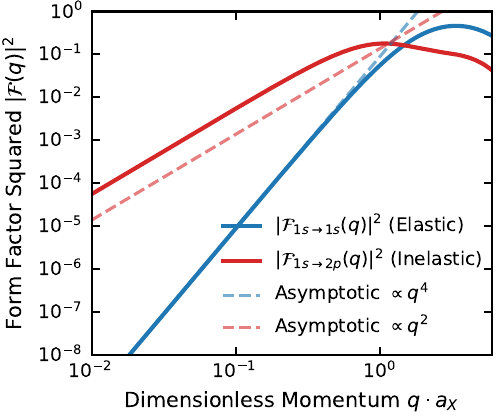}
\caption{\label{fig:form_factors}
Excitonic form factors $|\Fcal(q)|^2$ versus dimensionless momentum
$q \cdot \aX$ for (a) the elastic ($1s \to 1s$) and (b) the
inelastic ($1s \to 2p$) channels at $\eta = 0.6$. The dashed lines
show the asymptotic power-law limits derived from the multipole
expansion. The elastic channel vanishes as $q^4$, while the
inelastic channel vanishes as $q^2$, confirming the parity
selection rule.
}
\end{figure}

Figure~\ref{fig:form_factors} displays the exact form factors
$|\Fcal(q)|^2$ as a function of the dimensionless momentum
$q \cdot \aX$ for both the elastic ($1s \to 1s$) and inelastic
($1s \to 2p$) channels, using a representative mass asymmetry of
$\eta = 0.6$ ($\alphae = 0.2$, $\alphah = 0.8$). In the
long-wavelength limit ($q \to 0$), the elastic channel scales as
$q^4$ in the form factor (corresponding to $q^2$ in the effective
coupling $|g_X|^2$), while the inelastic channel scales as $q^2$ in
the form factor (corresponding to $q^0$ in $|g_X|^2$). These scaling
laws are in exact agreement with the multipole expansion.

To quantify the agreement, we extract the leading-order coefficients
via numerical differentiation at $q \cdot \aX = 0.001$. The results
are summarized in Table~\ref{tab:coefficients}.

\begin{table}[b]
\centering
\caption{\label{tab:coefficients}
Leading-order asymptotic coefficients of the excitonic form factor
$\Fcal(q)$ at low momentum transfer. Numerical values are obtained
from exact integration of the hydrogenic wavefunctions; analytical
values are derived from the Taylor expansion in
Eq.~\eqref{eq:taylor}.
}
\begin{tabular}{@{}lccc@{}}
\toprule
Transition & Order & Numerical & Analytical \\
\midrule
$1s \to 1s$ (elastic) & $q^2$ & $0.300000$ &
$\tfrac{1}{2}\eta\langle z^2\rangle = 0.30$ \\
$1s \to 2p$ (inelastic) & $q^1$ & $0.744934$ &
$\langle 2p_z|z|1s\rangle = 2^7\sqrt{2}/3^5$ \\
\bottomrule
\end{tabular}
\end{table}

The exact agreement between the numerical and analytical coefficients
for the elastic channel confirms that the multipole expansion captures
the complete low-$q$ physics. The elastic coefficient $0.30$ follows analytically: for $\eta=0.6$ the
quadrupolar prefactor is $\tfrac12\eta=0.3$, and with
$\langle 1s|z^2|1s\rangle=a_X^2$ (since $\langle r^2\rangle=3a_X^2$)
one obtains $F(q)\approx0.3\,(qa_X)^2$, in exact agreement with
Table~\ref{tab:coefficients}.

\subsection{\label{sec:parity}Parity-Controlled Interference: The
Selection Rule}

\begin{figure}[b]
\centering
\includegraphics[width=\columnwidth]{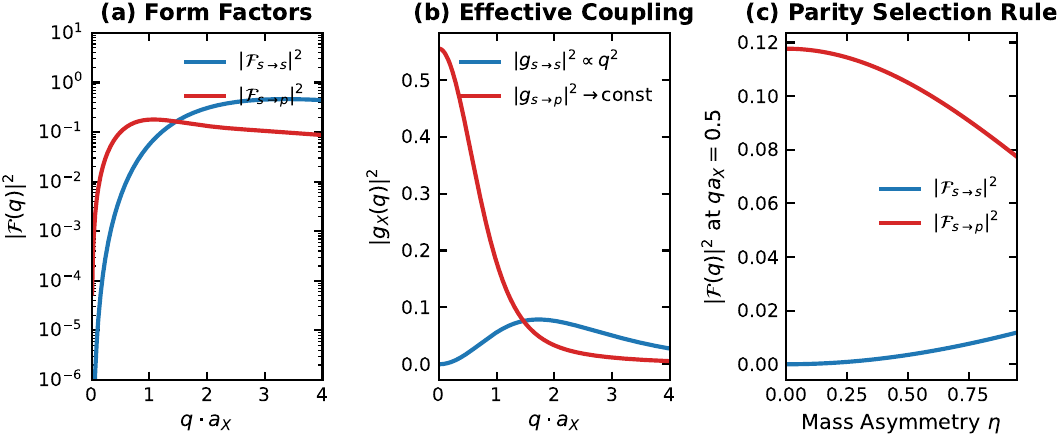}
\caption{\label{fig:parity}
Parity selection rule. (a) Form factors $|\Fcal(q)|^2$ versus $q
\cdot \aX$ for the elastic (blue) and inelastic (red) channels at
$\eta = 0.6$. (b) Effective Fr\"ohlich coupling $|g_X(q)|^2$ showing
the cancellation of the $1/q$ divergence for the elastic channel.
(c) Form factors versus mass asymmetry $\eta$ at fixed $q \cdot \aX
= 0.5$, demonstrating destructive interference for same-parity
transitions (blue, vanishing at $\eta = 0$) and constructive
interference for opposite-parity transitions (red, maximal at
$\eta = 0$).
}
\end{figure}

The most striking consequence of the multipole expansion is the
parity selection rule. We evaluate the form factors at a fixed
momentum transfer $q \cdot \aX = 0.5$ across the full range of mass
asymmetries $\eta \in [0, 0.95]$. The results are presented in
Table~\ref{tab:parity} and Figure~\ref{fig:parity}.

\begin{table}[b]
\centering
\caption{\label{tab:parity}
Parity-controlled interference at fixed $q \cdot \aX = 0.5$. The
ratio $|\Fcal_{sp}|^2 / |\Fcal_{ss}|^2$ quantifies the dominance of
the constructive channel over the destructive channel.
}
\begin{tabular}{@{}ccccc@{}}
\toprule
$\eta$ & $\alphae$ & $\alphah$ & $|\Fcal_{ss}|^2$ & $|\Fcal_{sp}|^2$ \\
\midrule
$0.00$ & $0.500$ & $0.500$ & $0.000000$ & $0.117702$ \\
$0.20$ & $0.400$ & $0.600$ & $0.000567$ & $0.115593$ \\
$0.40$ & $0.300$ & $0.700$ & $0.002246$ & $0.109458$ \\
$0.60$ & $0.200$ & $0.800$ & $0.004965$ & $0.099849$ \\
$0.80$ & $0.100$ & $0.900$ & $0.008612$ & $0.087601$ \\
\bottomrule
\end{tabular}
\end{table}

At $\eta = 0$ (equal electron and hole masses), the elastic form
factor vanishes to machine precision:
$|\Fcal_{ss}|^2 < 10^{-16}$. This is not an approximation; it is an
exact consequence of the parity symmetry of the $1s$ envelope. The
exciton is perfectly shielded from long-wavelength polar phonon
scattering. Simultaneously, the inelastic channel reaches its maximum
value, $|\Fcal_{sp}|^2 = 0.1177$, demonstrating that equal masses
produce perfect constructive interference for opposite-parity
transitions.

As the mass asymmetry increases, the elastic channel is progressively
activated through the quadrupolar term $\propto \eta$, while the
inelastic channel remains remarkably robust, decreasing by less than
26\% across the entire range $\eta \in [0, 0.8]$. This asymmetry in
sensitivity is the hallmark of the parity selection rule: the
destructive interference of the elastic channel is fragile and easily
broken by mass asymmetry, while the constructive interference of the
inelastic channel is topologically protected by the parity mismatch.

\subsection{\label{sec:phase}Integrated Polaronic Dressing: The Phase
Diagram}

The total exciton--phonon dressing strength is obtained by integrating
the squared matrix element over the 3D Brillouin zone. The integration
measure in three dimensions is $d^3q = 4\pi q^2\, dq$. Since the
Fr\"ohlich potential contributes a factor of $1/q^2$, the integrand
simplifies dramatically:
\begin{equation}
\label{eq:lambda_integral}
\lambda_X \propto \int_0^{\qmax}
\frac{|\Fcal(q)|^2}{q^2}\, q^2\, dq
= \int_0^{\qmax} |\Fcal(q)|^2\, dq.
\end{equation}
The $q^2$ density-of-states factor exactly cancels the $1/q^2$
Fr\"ohlich divergence. The total dressing parameter $\lambda_X$ is
therefore governed entirely by the internal form factor, integrated up
to the Brillouin zone boundary. We parameterize the integration cutoff
by the dimensionless exciton size
$\Lambda = \qmax\, \aX \approx \pi\, \aX / \alat$, which measures
the exciton Bohr radius in units of the lattice constant.

Figure~\ref{fig:phase} presents the integrated dressing parameters
$\lambda_{ss}$ and $\lambda_{sp}$ as functions of $\Lambda$ and
$\eta$. The phase diagram reveals three distinct regimes:

\begin{figure}[b]
\centering
\includegraphics[width=\columnwidth]{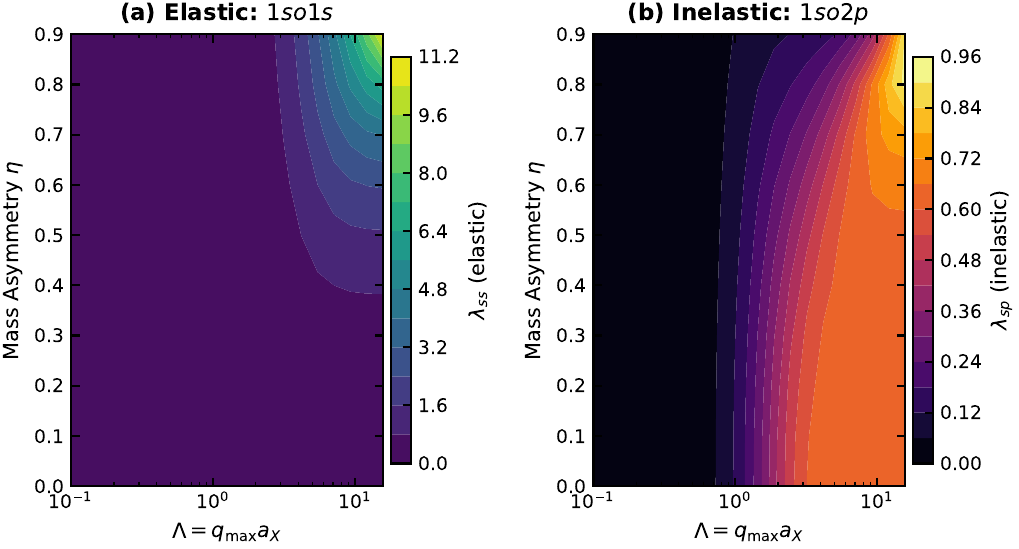}
\caption{\label{fig:phase}
Integrated polaronic dressing phase diagram. (a) Elastic channel
($1s \to 1s$): the dressing $\lambda_{ss}$ is identically zero at
$\eta = 0$ and grows with mass asymmetry. (b) Inelastic channel
($1s \to 2p$): the dressing $\lambda_{sp} \approx 0.64$ is
essentially independent of mass asymmetry, demonstrating the
robustness of constructive interference. The dimensionless exciton
size is $\Lambda = \qmax \aX$.
}
\end{figure}

\begin{table}[b]
\centering
\caption{\label{tab:phase}
Integrated dressing parameters in the large-exciton limit
($\Lambda = 15.8$).
}
\begin{tabular}{@{}cccc@{}}
\toprule
$\eta$ & $\lambda_{ss}$ (elastic) & $\lambda_{sp}$ (inelastic) &
$\lambda_{sp}/\lambda_{ss}$ \\
\midrule
$0.00$ & $0.0000$ & $0.6435$ & $\infty$ \\
$0.20$ & $0.0475$ & $0.6422$ & $13.5$ \\
$0.40$ & $0.1949$ & $0.6388$ & $3.3$ \\
$0.60$ & $0.4587$ & $0.6351$ & $1.4$ \\
$0.80$ & $0.8724$ & $0.6348$ & $0.7$ \\
\bottomrule
\end{tabular}
\end{table}

\textbf{Regime I: Perfect Protection} ($\eta = 0$, any $\Lambda$).
When the electron and hole masses are equal, the elastic dressing
$\lambda_{ss}$ is identically zero for all exciton sizes. The exciton
is completely decoupled from the long-range polar field in the elastic
channel. This regime is realized approximately in materials with
balanced band structures, such as certain halide perovskites.

\textbf{Regime II: Partial Protection}
($0 < \eta < 0.5$, large $\Lambda$). Finite mass asymmetry activates
the quadrupolar elastic channel, but the dressing remains
parametrically small compared to the inelastic channel. For
$\eta = 0.4$ and $\Lambda \gg 1$, the elastic dressing is
$\lambda_{ss} \approx 0.19$, roughly a factor of 3 smaller than the
inelastic dressing $\lambda_{sp} \approx 0.64$.

\textbf{Regime III: Broken Protection}
($\eta > 0.5$, large $\Lambda$). For strongly asymmetric systems,
the elastic dressing becomes comparable to the inelastic dressing. At
$\eta = 0.8$, $\lambda_{ss} \approx 0.87$ while
$\lambda_{sp} \approx 0.63$. In this regime, the quadrupolar channel
dominates, and the exciton is significantly dressed even in the
elastic scattering process.

A critical observation from Table~\ref{tab:phase} is that the
inelastic dressing $\lambda_{sp}$ is essentially constant
($\approx 0.64$) across all mass asymmetries. This robustness arises
because the dipole matrix element $\langle 2p | z | 1s \rangle$
depends only on the sum $\alphae + \alphah = 1$, not on the
individual mass fractions. The constructive interference channel is
therefore a fundamental feature of any excitonic system with broken
inversion symmetry, independent of the specific band structure.

\subsection{\label{sec:materials}Connection to Real Materials}

While our analysis is based on a minimal hydrogenic model, the
physical mechanisms identified here are broadly applicable and directly
applicable to a wide range of material systems. We briefly discuss
three representative cases.

\textbf{Halide perovskites (e.g., CsPbBr$_3$):} These materials
exhibit relatively balanced electron and hole effective masses
($m_e \approx 0.14\,m_0$, $m_h \approx 0.15\,m_0$, giving
$\eta \approx 0.03$), placing them squarely in Regime
I~\cite{even_perovskite,sendner_perovskite}. The parity selection
rule predicts that the $1s$ exciton in these materials should be
strongly protected against elastic Fr\"ohlich scattering, consistent
with the anomalously narrow excitonic linewidths observed
experimentally.

\textbf{Transition-metal dichalcogenides (e.g., MoS$_2$):} The
moderate effective mass asymmetry in monolayer TMDs
($m_e \approx 0.35\,m_0$, $m_h \approx 0.45\,m_0$,
$\eta \approx 0.13$) places them in Regime
II~\cite{kormanyos_tmd,chernikov_tmd}. The small exciton Bohr radius
($\aX \approx 10$~\AA) yields $\Lambda \approx 10$, where the
elastic dressing is finite but parametrically suppressed relative to
the inelastic channel. This is consistent with the observation that
the $1s$ exciton in MoS$_2$ exhibits a sharp zero-phonon line
accompanied by pronounced phonon sidebands at finite temperature.

\textbf{Rutile TiO$_2$:} TiO$_2$ represents a strongly asymmetric
system ($m_e \approx 0.6\,m_0$, $m_h \approx 2.5\,m_0$,
$\eta \approx 0.61$), placing it in Regime III where the elastic
quadrupolar channel is fully activated~\cite{tang_tio2}. The parity
protection of the $1s$ exciton is substantially broken, and both
elastic and inelastic channels contribute comparably to the total
polaronic dressing. This is consistent with the strong
electron--phonon coupling observed in TiO$_2$ and the significant
temperature-dependent renormalization of its optical gap.

\section{\label{sec:2d}Extension to Two-Dimensional Materials}

While the derivation in Sec.~\ref{sec:model} and the numerical phase
diagram in Sec.~\ref{sec:results} were formulated for
three-dimensional bulk semiconductors, the physical mechanisms
identified here generalize seamlessly to two-dimensional (2D) systems.
The extension to 2D is particularly relevant for polar monolayers,
such as TMDs, and for van der Waals heterostructures, where
exciton--phonon coupling dictates both low-temperature linewidths and
high-temperature ionization rates.

\subsection{\label{sec:2d_parity}Dimensional Invariance of the Parity
Selection Rule}

In a strictly 2D semiconductor, the exciton envelope functions
$\Phi_{n}(\mathbf{r})$ are solutions to the 2D hydrogenic
Schr\"odinger equation. The in-plane momentum transfer $\mathbf{q}$
is a 2D vector, and the phase operator in the excitonic form factor
remains $e^{i\mathbf{q}\cdot\mathbf{r}}$.

The Taylor expansion of this operator in 2D yields:
\begin{equation}
\label{eq:2d_taylor}
e^{i\mathbf{q}\cdot\mathbf{r}}
= 1 + i(\mathbf{q}\cdot\mathbf{r})
- \frac{1}{2}(\mathbf{q}\cdot\mathbf{r})^2
+ \mathcal{O}(q^3).
\end{equation}
Because the 2D ground-state envelope $|1s\rangle$ possesses circular
symmetry (and is therefore even under spatial inversion
$\mathbf{r} \to -\mathbf{r}$), the dipole expectation value
$\langle 1s | \mathbf{r} | 1s \rangle$ vanishes identically, just as
in 3D. The leading non-vanishing term for the elastic ($1s \to 1s$)
channel is again the quadrupole, scaling as $\mathcal{O}(q^2)$.
Conversely, the transition to the 2D $|2p\rangle$ state (which has
odd parity) allows the dipole term to survive, yielding a form factor
scaling as $\mathcal{O}(q^1)$.

Therefore, the parity-controlled destructive and constructive
interference mechanisms derived in Sec.~\ref{sec:multipole} are
\emph{dimensionally invariant}. The exciton's internal spatial
symmetry dictates the low-$q$ asymptotic limits, regardless of the
dimensionality of the host crystal.

\subsection{\label{sec:2d_frohlich}Modified Infrared Limits in 2D}

While the scaling of the form factor $|\Fcal(q)|^2$ is identical in
2D and 3D, the integration over the Brillouin zone and the exact form
of the Fr\"ohlich potential are modified by reduced dimensionality.
In 2D, the momentum integration measure is $d^2q = 2\pi q\, dq$.

For polar 2D materials (such as hBN or monolayer TMDs), the
long-range macroscopic electric field generated by in-plane
longitudinal optical (LO) phonons yields a bare Fr\"ohlich potential
that scales as $V_{2D}(q) \propto 1/q$~\cite{sohier_2d}.
Consequently, the squared matrix element scales as
$|V_{2D}(q)|^2 \propto 1/q^2$.

The effective exciton--phonon coupling squared is therefore:
\begin{equation}
\label{eq:2d_g_squared}
|g_X^{2D}(q)|^2
= |V_{2D}(q)|^2\, |\Fcal(q)|^2
\propto \frac{1}{q^2}\, |\Fcal(q)|^2.
\end{equation}
Applying our dimensionally invariant form factor scaling:
\begin{align}
\text{Elastic } (1s \to 1s): &\quad
|g_X^{2D}(q)|^2 \propto \frac{1}{q^2} \times q^4 = q^2,
\label{eq:2d_elastic}\\
\text{Inelastic } (1s \to 2p): &\quad
|g_X^{2D}(q)|^2 \propto \frac{1}{q^2} \times q^2 = q^0.
\label{eq:2d_inelastic}
\end{align}
Just as in 3D, the elastic coupling vanishes as $q \to 0$, while the
inelastic coupling remains finite. However, the total integrated
dressing parameter
$\lambda_X^{2D} \propto \int |g_X^{2D}(q)|^2\, q\, dq$
reveals a distinct 2D suppression:
\begin{align}
\lambda_{ss}^{2D} &\propto \int_0^{\qmax} (q^2)\, q\, dq
\propto \qmax^4, \label{eq:2d_lambda_ss}\\
\lambda_{sp}^{2D} &\propto \int_0^{\qmax} (q^0)\, q\, dq
\propto \qmax^2. \label{eq:2d_lambda_sp}
\end{align}
Both integrals converge smoothly to zero at the band edge. However,
the inelastic dressing is larger than the elastic dressing by a factor
of $1/q^2$. This mathematical hierarchy provides a rigorous
explanation for experimental observations in TMDs: the ground-state
$1s$ exciton exhibits anomalously narrow zero-phonon linewidths
(protected by elastic destructive interference), while the optical
absorption and photoluminescence spectra are dominated by strong,
broad phonon sidebands driven by the $1s \to 2p$ (and $1s \to$
continuum) constructive inelastic channels.

\subsection{\label{sec:interlayer}Breaking Protection: Interlayer
Excitons}

The parity selection rule assumes that the electron and the hole
couple to the \emph{same} macroscopic phonon field and share a
spatial envelope with definite parity. This protection is dramatically
broken in \textbf{interlayer excitons} formed in type-II van der
Waals heterostructures (e.g., MoSe$_2$/WSe$_2$).

In an interlayer exciton, the electron and hole are confined to
different monolayers separated by a vertical distance $d$. This
spatial separation breaks the parity selection rule in two fundamental
ways:

\begin{enumerate}
\item \textbf{Out-of-plane coupling:} For phonon modes with a finite
out-of-plane momentum component $q_z$, the phase difference between
the electron and hole is $e^{i q_z d} - 1 \approx i q_z d$. The
dipole term survives even for the $1s \to 1s$ ground state, yielding
a form factor $\Fcal(q) \propto \mathcal{O}(q^0)$. The exciton
acquires a permanent out-of-plane dipole moment and couples strongly
to out-of-plane polar fields.

\item \textbf{Decoupled phonon baths:} If the interlayer spacing is
sufficiently large, the electron couples primarily to the LO phonons
of the acceptor layer, while the hole couples to the LO phonons of
the donor layer. The interaction Hamiltonian becomes
$\Heph \propto g_1 \hat{\rho}_e \hat{b}_1
- g_2 \hat{\rho}_h \hat{b}_2$. Because the electron and hole are
scattering off \emph{distinct phonon operators}, the destructive
interference term $\langle \hat{b}_1 \hat{b}_2^\dagger \rangle$
vanishes. The cross-cancellation is physically severed, and both
elastic and inelastic channels exhibit unsuppressed, macroscopic
Fr\"ohlich coupling.
\end{enumerate}

This framework naturally unifies the seemingly contradictory
experimental reports of weak versus strong phonon dressing in 2D
materials: intralayer excitons are protected by parity, while
interlayer excitons are inherently vulnerable to polaronic dressing.

\subsection{\label{sec:ste}Charge-Transfer Excitons and Self-Trapping}
The parity selection rule derived in Sec.\ \ref{sec:multipole} fundamentally relies on the spatial overlap of the electron and hole envelope functions. When this overlap is broken, the destructive interference is severed. A recent experimental realization of this occurs in charge-transfer excitons, such as those in lead-free double halide perovskites (e.g., Cs$_2$Ag$_{0.4}$Na$_{0.6}$InCl$_6$)~\cite{xu_ste_cpl_2026}. In these materials, photoexcitation induces charge transfer such that electrons and holes localize on distinct atomic sites (e.g., In and Ag octahedra, respectively). This spatial and orbital separation ensures that the electron-phonon and hole-phonon deformation potentials cannot cancel. The resulting unsuppressed exciton-phonon coupling drives massive lattice distortions, binding the exciton to form a self-trapped exciton (STE) or exciton-polaron~\cite{xu_ste_cpl_2026}. Recent experiments have further revealed Fano resonances in these systems, highlighting the complex many-body quantum interference between discrete phonon modes and continuum exciton states that emerges once the primary electron-hole cancellation is broken~\cite{xu_ste_cpl_2026}.

\section{\label{sec:conclusions}Conclusions}

We have demonstrated that exciton--phonon coupling in polar
semiconductors is governed by a parity-controlled interference
selection rule. By performing an exact multipole expansion of the
excitonic form factor and validating it against numerical integration
of hydrogenic envelope functions, we have established the following
key results:

\begin{enumerate}
\item The long-range $1/q$ infrared divergence of the Fr\"ohlich
interaction is exactly canceled for elastic exciton--phonon
scattering between states of the same parity. The ground-state $1s$
exciton is protected from long-wavelength polar phonons by
destructive electron--hole interference.

\item This protection is fragile: finite mass asymmetry between the
electron and hole activates higher-order multipole terms, with the
leading correction scaling as $\eta\, q^2$. The elastic dressing
grows quadratically with mass asymmetry.

\item In contrast, inelastic scattering channels connecting states of
opposite parity exhibit constructive interference. The coupling
remains finite and robust in the long-wavelength limit, independent
of mass asymmetry. This constructive channel is a (within the long-wavelength effective-mass Fr\"ohlich framework) universal feature
of any excitonic system with broken parity.

\item The total integrated dressing parameter $\lambda_X$ exhibits a
clean phase structure as a function of exciton size $\Lambda$ and
mass asymmetry $\eta$, with three distinct regimes: perfect
protection, partial protection, and broken protection.

\item The selection rule is dimensionally invariant, applying equally
to bulk and two-dimensional systems. In 2D, the modified density of
states further suppresses the elastic channel relative to the
inelastic channel.
\end{enumerate}

These results provide a computationally inexpensive, analytically
exact framework for understanding and predicting exciton--phonon
interactions across a wide range of material systems, from bulk
semiconductors to 2D materials and van der Waals heterostructures.
The parity selection rule offers a new design principle for
engineering excitonic materials with tailored phonon coupling: by
controlling the mass asymmetry and the parity of the excitonic
states, one can tune the exciton--phonon interaction from fully
protected to strongly dressed.

\begin{acknowledgments}
The author acknowledges Kenya Education Network (KENET) research services for computing resources.
\end{acknowledgments}

\bibliography{references}

\begin{thebibliography}{12}%
\makeatletter
\providecommand \@ifxundefined [1]{%
 \@ifx{#1\undefined}
}%
\providecommand \@ifnum [1]{%
 \ifnum #1\expandafter \@firstoftwo
 \else \expandafter \@secondoftwo
 \fi
}%
\providecommand \@ifx [1]{%
 \ifx #1\expandafter \@firstoftwo
 \else \expandafter \@secondoftwo
 \fi
}%
\providecommand \natexlab [1]{#1}%
\providecommand \enquote  [1]{``#1''}%
\providecommand \bibnamefont  [1]{#1}%
\providecommand \bibfnamefont [1]{#1}%
\providecommand \citenamefont [1]{#1}%
\providecommand \href@noop [0]{\@secondoftwo}%
\providecommand \href [0]{\begingroup \@sanitize@url \@href}%
\providecommand \@href[1]{\@@startlink{#1}\@@href}%
\providecommand \@@href[1]{\endgroup#1\@@endlink}%
\providecommand \@sanitize@url [0]{\catcode `\\12\catcode `\$12\catcode `\&12\catcode `\#12\catcode `\^12\catcode `\_12\catcode `\%12\relax}%
\providecommand \@@startlink[1]{}%
\providecommand \@@endlink[0]{}%
\providecommand \url  [0]{\begingroup\@sanitize@url \@url }%
\providecommand \@url [1]{\endgroup\@href {#1}{\urlprefix }}%
\providecommand \urlprefix  [0]{URL }%
\providecommand \Eprint [0]{\href }%
\providecommand \doibase [0]{https://doi.org/}%
\providecommand \selectlanguage [0]{\@gobble}%
\providecommand \bibinfo  [0]{\@secondoftwo}%
\providecommand \bibfield  [0]{\@secondoftwo}%
\providecommand \translation [1]{[#1]}%
\providecommand \BibitemOpen [0]{}%
\providecommand \bibitemStop [0]{}%
\providecommand \bibitemNoStop [0]{.\EOS\space}%
\providecommand \EOS [0]{\spacefactor3000\relax}%
\providecommand \BibitemShut  [1]{\csname bibitem#1\endcsname}%
\let\auto@bib@innerbib\@empty
\bibitem [{\citenamefont {Giustino}(2017)}]{giustino_rmp_2017}%
  \BibitemOpen
  \bibfield  {author} {\bibinfo {author} {\bibfnamefont {F.}~\bibnamefont {Giustino}},\ }\bibfield  {title} {\bibinfo {title} {Electron-phonon interactions from first principles},\ }\href {https://doi.org/10.1103/RevModPhys.89.015003} {\bibfield  {journal} {\bibinfo  {journal} {Rev. Mod. Phys.}\ }\textbf {\bibinfo {volume} {89}},\ \bibinfo {pages} {015003} (\bibinfo {year} {2017})}\BibitemShut {NoStop}%
\bibitem [{\citenamefont {Diab}\ \emph {et~al.}(2016)\citenamefont {Diab}, \citenamefont {Tripp{\'{e}}-Allard}, \citenamefont {L{\'{e}}d{\'{e}}e}, \citenamefont {Jemli}, \citenamefont {Vilar}, \citenamefont {Bouchez}, \citenamefont {Jacques}, \citenamefont {Tejeda}, \citenamefont {Even}, \citenamefont {Lauret}, \citenamefont {Deleporte},\ and\ \citenamefont {Garrot}}]{Diab2016JPCLETT}%
  \BibitemOpen
  \bibfield  {author} {\bibinfo {author} {\bibfnamefont {H.}~\bibnamefont {Diab}}, \bibinfo {author} {\bibfnamefont {G.}~\bibnamefont {Tripp{\'{e}}-Allard}}, \bibinfo {author} {\bibfnamefont {F.}~\bibnamefont {L{\'{e}}d{\'{e}}e}}, \bibinfo {author} {\bibfnamefont {K.}~\bibnamefont {Jemli}}, \bibinfo {author} {\bibfnamefont {C.}~\bibnamefont {Vilar}}, \bibinfo {author} {\bibfnamefont {G.}~\bibnamefont {Bouchez}}, \bibinfo {author} {\bibfnamefont {V.~L.~R.}\ \bibnamefont {Jacques}}, \bibinfo {author} {\bibfnamefont {A.}~\bibnamefont {Tejeda}}, \bibinfo {author} {\bibfnamefont {J.}~\bibnamefont {Even}}, \bibinfo {author} {\bibfnamefont {J.-S.}\ \bibnamefont {Lauret}}, \bibinfo {author} {\bibfnamefont {E.}~\bibnamefont {Deleporte}},\ and\ \bibinfo {author} {\bibfnamefont {D.}~\bibnamefont {Garrot}},\ }\bibfield  {title} {\bibinfo {title} {Narrow linewidth excitonic emission in organic--inorganic lead iodide perovskite single crystals},\ }\href {https://doi.org/10.1021/acs.jpclett.6b02261} {\bibfield  {journal}
  {\bibinfo  {journal} {The Journal of Physical Chemistry Letters}\ }\textbf {\bibinfo {volume} {7}},\ \bibinfo {pages} {5093} (\bibinfo {year} {2016})}\BibitemShut {NoStop}%
\bibitem [{\citenamefont {Wright}\ \emph {et~al.}(2016)\citenamefont {Wright}, \citenamefont {Verdi}, \citenamefont {Milot}, \citenamefont {Eperon}, \citenamefont {P{\'{e}}rez-Osorio}, \citenamefont {Snaith}, \citenamefont {Giustino},\ and\ \citenamefont {Johnston}}]{Wright2016}%
  \BibitemOpen
  \bibfield  {author} {\bibinfo {author} {\bibfnamefont {A.~D.}\ \bibnamefont {Wright}}, \bibinfo {author} {\bibfnamefont {C.}~\bibnamefont {Verdi}}, \bibinfo {author} {\bibfnamefont {R.~L.}\ \bibnamefont {Milot}}, \bibinfo {author} {\bibfnamefont {G.~E.}\ \bibnamefont {Eperon}}, \bibinfo {author} {\bibfnamefont {M.~A.}\ \bibnamefont {P{\'{e}}rez-Osorio}}, \bibinfo {author} {\bibfnamefont {H.~J.}\ \bibnamefont {Snaith}}, \bibinfo {author} {\bibfnamefont {F.}~\bibnamefont {Giustino}},\ and\ \bibinfo {author} {\bibfnamefont {M.~B.}\ \bibnamefont {Johnston}},\ }\bibfield  {title} {\bibinfo {title} {Electron--phonon coupling in hybrid lead halide perovskites},\ }\href {https://doi.org/10.1038/ncomms11755} {\bibfield  {journal} {\bibinfo  {journal} {Nature Communications}\ }\textbf {\bibinfo {volume} {7}},\ \bibinfo {pages} {11755} (\bibinfo {year} {2016})}\BibitemShut {NoStop}%
\bibitem [{\citenamefont {Atambo}(2026{\natexlab{a}})}]{atambo_lattice_2026}%
  \BibitemOpen
  \bibfield  {author} {\bibinfo {author} {\bibfnamefont {M.~O.}\ \bibnamefont {Atambo}},\ }\href@noop {} {\bibinfo {title} {Symmetry-protected intraband exciton--phonon scattering and its breakdown by mass asymmetry}} (\bibinfo {year} {2026}{\natexlab{a}})\BibitemShut {NoStop}%
\bibitem [{\citenamefont {Atambo}(2026{\natexlab{b}})}]{atambo_recoil_2026}%
  \BibitemOpen
  \bibfield  {author} {\bibinfo {author} {\bibfnamefont {M.~O.}\ \bibnamefont {Atambo}},\ }\href@noop {} {\bibinfo {title} {Why polar excitons stay sharp: parity protection of the center-of-mass recoil channel in exciton--phonon scattering}} (\bibinfo {year} {2026}{\natexlab{b}})\BibitemShut {NoStop}%
\bibitem [{\citenamefont {Even}\ \emph {et~al.}(2013)\citenamefont {Even}, \citenamefont {Pedesseau}, \citenamefont {Jancu},\ and\ \citenamefont {Katan}}]{even_perovskite}%
  \BibitemOpen
  \bibfield  {author} {\bibinfo {author} {\bibfnamefont {J.}~\bibnamefont {Even}}, \bibinfo {author} {\bibfnamefont {L.}~\bibnamefont {Pedesseau}}, \bibinfo {author} {\bibfnamefont {J.-M.}\ \bibnamefont {Jancu}},\ and\ \bibinfo {author} {\bibfnamefont {C.}~\bibnamefont {Katan}},\ }\bibfield  {title} {\bibinfo {title} {Importance of spin-orbit coupling in hybrid organic/inorganic perovskites for photovoltaic applications},\ }\href {https://doi.org/10.1021/jz401532q} {\bibfield  {journal} {\bibinfo  {journal} {The Journal of Physical Chemistry Letters}\ }\textbf {\bibinfo {volume} {4}},\ \bibinfo {pages} {2999} (\bibinfo {year} {2013})}\BibitemShut {NoStop}%
\bibitem [{\citenamefont {Sendner}\ \emph {et~al.}(2016)\citenamefont {Sendner} \emph {et~al.}}]{sendner_perovskite}%
  \BibitemOpen
  \bibfield  {author} {\bibinfo {author} {\bibfnamefont {M.}~\bibnamefont {Sendner}} \emph {et~al.},\ }\bibfield  {title} {\bibinfo {title} {Optical phonons in methylammonium lead iodide and their coupling to the exciton},\ }\href@noop {} {\bibfield  {journal} {\bibinfo  {journal} {Mater. Horiz.}\ }\textbf {\bibinfo {volume} {3}},\ \bibinfo {pages} {613} (\bibinfo {year} {2016})}\BibitemShut {NoStop}%
\bibitem [{\citenamefont {Korm{\'{a}}nyos}\ \emph {et~al.}(2015)\citenamefont {Korm{\'{a}}nyos}, \citenamefont {Burkard}, \citenamefont {Gmitra}, \citenamefont {Fabian}, \citenamefont {Z{\'{o}}lyomi}, \citenamefont {Drummond},\ and\ \citenamefont {Fal'ko}}]{kormanyos_tmd}%
  \BibitemOpen
  \bibfield  {author} {\bibinfo {author} {\bibfnamefont {A.}~\bibnamefont {Korm{\'{a}}nyos}}, \bibinfo {author} {\bibfnamefont {G.}~\bibnamefont {Burkard}}, \bibinfo {author} {\bibfnamefont {M.}~\bibnamefont {Gmitra}}, \bibinfo {author} {\bibfnamefont {J.}~\bibnamefont {Fabian}}, \bibinfo {author} {\bibfnamefont {V.}~\bibnamefont {Z{\'{o}}lyomi}}, \bibinfo {author} {\bibfnamefont {N.~D.}\ \bibnamefont {Drummond}},\ and\ \bibinfo {author} {\bibfnamefont {V.~I.}\ \bibnamefont {Fal'ko}},\ }\bibfield  {title} {\bibinfo {title} {k$\cdot$p theory for two-dimensional transition metal dichalcogenide semiconductors},\ }\href {https://doi.org/10.1088/2053-1583/2/2/022001} {\bibfield  {journal} {\bibinfo  {journal} {2D Materials}\ }\textbf {\bibinfo {volume} {2}},\ \bibinfo {pages} {022001} (\bibinfo {year} {2015})}\BibitemShut {NoStop}%
\bibitem [{\citenamefont {Chernikov}\ \emph {et~al.}(2014)\citenamefont {Chernikov} \emph {et~al.}}]{chernikov_tmd}%
  \BibitemOpen
  \bibfield  {author} {\bibinfo {author} {\bibfnamefont {A.}~\bibnamefont {Chernikov}} \emph {et~al.},\ }\bibfield  {title} {\bibinfo {title} {Exciton binding energy and nonhydrogenic rydberg series in monolayer {WS$_2$}},\ }\href@noop {} {\bibfield  {journal} {\bibinfo  {journal} {Phys. Rev. Lett.}\ }\textbf {\bibinfo {volume} {113}},\ \bibinfo {pages} {076802} (\bibinfo {year} {2014})}\BibitemShut {NoStop}%
\bibitem [{\citenamefont {Tang}\ \emph {et~al.}(2012)\citenamefont {Tang} \emph {et~al.}}]{tang_tio2}%
  \BibitemOpen
  \bibfield  {author} {\bibinfo {author} {\bibfnamefont {H.}~\bibnamefont {Tang}} \emph {et~al.},\ }\bibfield  {title} {\bibinfo {title} {Electronic structure and optical properties of rutile {TiO$_2$} from first principles},\ }\href@noop {} {\bibfield  {journal} {\bibinfo  {journal} {Phys. Rev. B}\ }\textbf {\bibinfo {volume} {85}},\ \bibinfo {pages} {195119} (\bibinfo {year} {2012})}\BibitemShut {NoStop}%
\bibitem [{\citenamefont {Sohier}\ \emph {et~al.}(2014)\citenamefont {Sohier}, \citenamefont {Calandra},\ and\ \citenamefont {Mauri}}]{sohier_2d}%
  \BibitemOpen
  \bibfield  {author} {\bibinfo {author} {\bibfnamefont {T.}~\bibnamefont {Sohier}}, \bibinfo {author} {\bibfnamefont {M.}~\bibnamefont {Calandra}},\ and\ \bibinfo {author} {\bibfnamefont {F.}~\bibnamefont {Mauri}},\ }\bibfield  {title} {\bibinfo {title} {Macroscopic dielectric response and excitonic coupling in two-dimensional materials},\ }\href@noop {} {\bibfield  {journal} {\bibinfo  {journal} {Phys. Rev. B}\ }\textbf {\bibinfo {volume} {90}},\ \bibinfo {pages} {125414} (\bibinfo {year} {2014})}\BibitemShut {NoStop}%
\bibitem [{\citenamefont {Xu}\ \emph {et~al.}(2026)\citenamefont {Xu}, \citenamefont {Liu}, \citenamefont {Pang}, \citenamefont {Zhang}, \citenamefont {Wang}, \citenamefont {Chang}, \citenamefont {Luo}, \citenamefont {Tang}, \citenamefont {Xiong}, \citenamefont {Meng}, \citenamefont {Gao},\ and\ \citenamefont {Zhang}}]{xu_ste_cpl_2026}%
  \BibitemOpen
  \bibfield  {author} {\bibinfo {author} {\bibfnamefont {K.-X.}\ \bibnamefont {Xu}}, \bibinfo {author} {\bibfnamefont {X.-B.}\ \bibnamefont {Liu}}, \bibinfo {author} {\bibfnamefont {S.}~\bibnamefont {Pang}}, \bibinfo {author} {\bibfnamefont {Z.}~\bibnamefont {Zhang}}, \bibinfo {author} {\bibfnamefont {Y.}~\bibnamefont {Wang}}, \bibinfo {author} {\bibfnamefont {H.}~\bibnamefont {Chang}}, \bibinfo {author} {\bibfnamefont {J.}~\bibnamefont {Luo}}, \bibinfo {author} {\bibfnamefont {J.}~\bibnamefont {Tang}}, \bibinfo {author} {\bibfnamefont {Q.}~\bibnamefont {Xiong}}, \bibinfo {author} {\bibfnamefont {S.}~\bibnamefont {Meng}}, \bibinfo {author} {\bibfnamefont {S.}~\bibnamefont {Gao}},\ and\ \bibinfo {author} {\bibfnamefont {J.}~\bibnamefont {Zhang}},\ }\bibfield  {title} {\bibinfo {title} {Quantum interference and optical tuning of self-trapped exciton state in double halide perovskite},\ }\href {https://doi.org/10.1088/0256-307X/43/3/030703} {\bibfield  {journal} {\bibinfo  {journal} {Chinese Physics Letters}\
  }\textbf {\bibinfo {volume} {43}},\ \bibinfo {pages} {030703} (\bibinfo {year} {2026})}\BibitemShut {NoStop}%
\end{thebibliography}%

\end{document}